\documentclass[twocolumn,pre,showpacs,superscriptaddress]{revtex4}
\usepackage{graphicx}
\usepackage{amsmath}
\usepackage{amssymb}

\begin{document}

\title{Antiphase synchronization of two nonidentical pendulums}

\author{Il Gu Yi}
\affiliation{Department of Physics, Sungkyunkwan University, Suwon 440-746, Korea}
\author{Hyun Keun Lee}
\affiliation{BK21 Physics Research Division, Sungkyunkwan University, Suwon 440-746, Korea}
\author{Sung Hyun Jeon}
\affiliation{Department of Physics, Ajou University, Suwon 442-749, Korea}
\author{Beom Jun Kim}
\email[Corresponding author, E-mail: ]{beomjun@skku.edu}
\affiliation{BK21 Physics Research Division and Department of Energy Science, Sungkyunkwan University, Suwon 440-746, Korea}
%\affiliation{Department of Computational Biology, School of Computer Science
%and Communication, Royal Institute of Technology, 100 44 Stockholm, Sweden}

%\date{\today}

\begin{abstract}
We numerically study the synchronization of two nonidentical pendulum motions,
pivoting on a common movable frame in the point of view of the dynamic phase
transition. When the difference in the pendulum lengths is not too large, it is
shown that the system settles down into the dynamic state of the antiphase
synchronization with the phase difference $\pi$. We observe that there is a
bistable region where either the antiphase synchronized state or the
desynchronized state can be stabilized. We also find that there exists a
hysteresis effect around the dynamic phase transition as the length difference
is adiabatically changed.
\end{abstract}

%05.45.Xt	Synchronization; coupled oscillators

%\keywords{Synchronization, Huygens's clocks, bifurcation, hysteresis}

\maketitle

The phrase of {\it odd kind of sympathy} (in short, odd sympathy) was used by a
prominent Dutch mathematician and physicist Christiaan Huygens in order to
mention an interesting phenomenon observed for the two pendulum clocks attached on the
wall~\cite{Pikovsky}: Even after an intentional disturbance,
two pendulum clocks evolve into a state in which they swing together in the
opposite directions in synchrony.
This is one of the most historical observations which naturally lead to the concept of
synchronization of dynamic variables, and have attracted
%The important point therein is that
%these variables were not designed to do so from the beginning. Hereby, the
%synchronization is a nontrivial phenomenon, and thus attracts
many scientific/engineering researchers for a long time~\cite{Pikovsky,Blekhman}. In the language of
synchronization study, the odd sympathy corresponds to the antiphase synchronization
in which the frequencies of the pendulums' oscillations become identical but the phases
show the mismatch $\pi$.

There are abundant examples of synchronization phenomena
from biological objects~\cite{SynchInBiology} and celestial systems~\cite{SynchInCelestial}
in nature to manmade electrical and mechanical systems~\cite{SynchInManMadeSystem}.
Interestingly, such synchronization behaviors have been in most cases empirically
discovered by chance {\it a posteriori} (see the Introduction of Ref.~\onlinecite{Blekhman}),
%~\cite{SynchInEmpiricalFounding},
and we still need to understand more to predict why and when it happens.
The lack of the precise knowledge can result in a noncontrollable outcome
or a catastrophic disaster in the worst case~\cite{BridgeCollapse-wiki}.
Even for such a simple setup of two pendulums, which allowed a chance for C. Huygens to
become aware of the odd sympathy, investigations have still been performed to
unveil the synchronization property in
it~\cite{Blekhman,Bennett,Pantaleon,Oud,Senator,Alexander}.
Such a study on the prototypical setup is necessary for a comprehensive
understanding of the synchronization phenomenon as a dynamic phase.
This is also to contribute to making the abstract-model-based synchronization
research more fruitful in the viewpoint of the synchronization
as a thermodynamic phase~\cite{freq}.%~\cite{AbsSynch}.

More than two hundred years later after the Huygens' observation of the odd
sympathy, a qualitative understanding  was firstly tried by Korteweg in
1906~\cite{Bennett}. Nearly one hundred more years later, in early 2000s,
Bennett {\it et. al.} revisited this issue~\cite{Bennett} and successfully
reproduced the antiphase synchronization in their experiment and
theoretical  model study. Other researchers of the two-pendulum system have
also been interested in the inphase synchronization, in which the frequencies
of the pendulums' oscillations are identical without the phase mismatch. The
possibility of the existence of an inphase state was suggested in
Ref.~\onlinecite{Blekhman}, and was credibly reproduced for the first time in
the experiment introduced in Ref.~\onlinecite{Pantaleon}. These works have been
followed by a series of studies where various experimental setups and mechanical
models have been proposed to understand the onset of inphase synchronization as
well as antiphase one~\cite{Oud,Senator,Alexander}. 

The motivation of this work is to investigate 
how generic the synchronization is in the two
pendulum system. This is attributed to the fact that each work
above~\cite{Bennett,Pantaleon,Oud,Senator,Alexander} is based on its own
specific experimental setup, for example, escapement mechanism and system
structure. We are also interested in the bifurcation property of the
synchronization state.
In this paper, we numerically study the antiphase synchronization of the two
nonidentical pendulums in the viewpoint of the dynamic phase transition.  
For the curiosity of generic feature, we introduce a model whose detail is as
simple as possible.
As the lengths of pendulums are varied, it is observed that there exists a broad
range of bistable region, where the antiphase synchronized state and the
desynchronized state coexist depending on initial conditions. The bistability
is also shown to lead to  a hysteresis effect as the lengths of  pendulums are
varied adiabatically.

\begin{figure}
\includegraphics[width=0.35\textwidth]{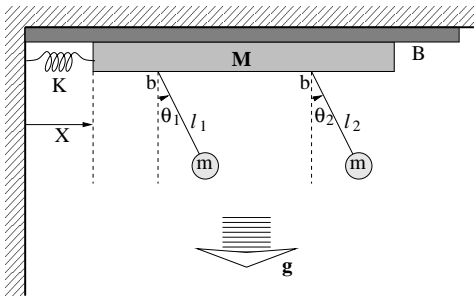}
\caption{Each pendulum ($i=1$ and 2) of the mass $m$ is connected to the rigid
common frame of the mass $M$ by the massless rigid rod of the length $l_i$.
The frame at the horizontal position $X$
is attached to a spring with the spring constant $K$. It
is assumed that the sliding motion of the frame and the pivot motion of the
pendulum is dissipative, which are described by the friction constant $B$ and
$b$, respectively.  The hatched region stands for the rigid immobile wall
considered as the reference frame and the uniform gravitational field ${\bf g}$
is applied vertically. The three movable objects (the frame and the two
pendulums) are described by the coordinates $X$, $\theta_1$, and $\theta_2$,
respectively, which are the three degrees of freedom in the system.}
\label{schematics}
\end{figure}

We first introduce the three degrees of freedom model~\cite{Bennett},
schematically shown in Fig.~\ref{schematics}.  We apply the Lagrangian least
action principle~\cite{goldstein} for nonconservative forces, i.e., the
escapement and the friction forces in this work, and achieve the equations of motion in
dimensionless form (see Fig.\ref{schematics} and compare with
Ref.~\onlinecite{Bennett})
\begin{eqnarray}
\label{dleom-a}
{\ddot \theta_{i}} + 2\gamma {\dot \theta_{i}} +
(\sin\theta_i + {\ddot x} \cos\theta_{i})/{\bar l}_i - f_i = 0 ,  \\
{\ddot x} + 2 \Gamma {\dot x} + \Omega^2 x
 + \mu \sum_i {\bar l}_i ({\ddot \theta_i} \cos\theta_{i} - {\dot \theta}^2_i \sin \theta_{i}) = 0,
\end{eqnarray}
where the frame coordinate $x (\equiv X/l)$ and the dimensionless length
${\bar l}_i (\equiv l_i/l)$ are measured in units of $l$ 
(${\bar l}_i = 1 + \epsilon_i$, where $\epsilon_i$ can be interpreted as
a small but unavoidable relative error in manufacturing or measurement), and  the time $t$ (the dots on the symbols represent time derivatives)
is in units of $\sqrt{l/g}$, respectively.
We have also defined the reduced mass as $\mu \equiv m/(M+2m)$,
the effective coupling strength of the frame as $\Omega^{2} \equiv K/(M+2m)$,
and the dimensionless friction coefficients for the frame
$\Gamma \equiv (B/2)(\sqrt{l/g})/(M+2m)$ and for the pendulum
$\gamma \equiv (b/2)\sqrt{l/g}$, respectively.
Since every degree of freedom in the system is subject to the damping,
the motion will eventually stop at $x=\theta_1 = \theta_2 = 0$ in the
absence of the external energy source.
As the escapement method, we apply the impulsive force $f_i$ in
Eq.~(\ref{dleom-a}) when $\dot\theta_i = 0$  [$f_i < 0  $ for $\theta_i > 0$ and
$f_i > 0  $ for $\theta_i < 0$].
Our escapement method enforces the pendulums not to stop by injecting kinetic
energy into the system. 
It should be noted that the death phase observed in Ref.~\onlinecite{Bennett}
is not allowed in our setup. Although other escapement mechanisms could
lead to different results,
we believe that our escapement algorithm is not too far from the reality,
and that the generic qualitative results obtained in this work should be
observable in properly prepared real experiments.
We also remark that the other mechanical detail such as friction in joints brings
about a complex phenomenon, which has been thoroughly studied in Ref.~\onlinecite{Awrejcewicz})

In the numerical experiments, we use $\gamma=3\times 10^{-4},\Omega=0,\Gamma=0.8$, and
$|f_1|=|f_2|=0.35$. We have tested the reduced mass $\mu=0.02, 0.025$, and $0.03$,
only to find insignificant differences, and the results presented in this
work are for $\mu = 0.025$.
We use the
4th-order Runge-Kutta algorithm to integrate equations of motion with
the discrete time step size $\Delta t = 0.01$.

\begin{figure}
\includegraphics[width=0.45\textwidth]{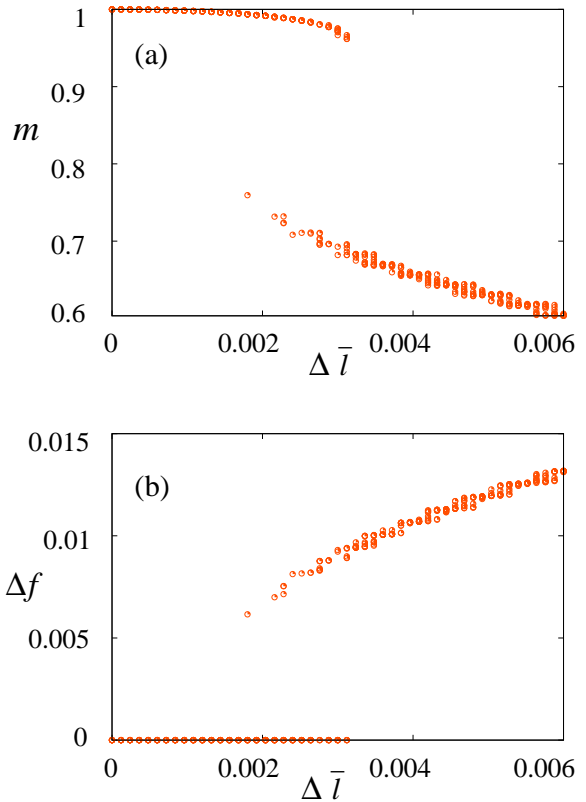}
\caption{(Color online)
(a) The phase synchronization order parameter $m$ and (b) the
frequency entrainment order parameter $\Delta f$ are shown.
The dimensionless length difference $\Delta {\bar l} = | {\bar l_1} - {\bar l_2} |$ 
is used for the horizontal axis with ${\bar l_1} = 1$, and each point
is obtained from the random initial condition with $\dot\theta_i = 0$ 
and $\theta_i \in \pm [0.05, 0.1]$. 
Clearly observed is the existences of two distinct dynamic phases,
the antiphase synchronized state and the desynchronized state: The former
is characterized by $m \approx 1$ and $\Delta f \approx 0$, while the latter
by  $m$  significantly less than unity and $\Delta f$ far from zero. It is to be
noted that there is a broad range of bistability where whether the system
settles down to the synchronized state or not depends on actual values of
$\bar{l}_1$ and $\bar{l}_2$. Change of initial conditions
is found to alter neither the dynamic phase transition point
nor the observed bistability significantly.
}
\label{bistability}
\end{figure}

In order to measure the degree of synchrony, we define the order parameter
$m$ for the phase synchronization as
\begin{equation}
m \equiv \left\langle  \cos \left( \Delta \phi - \pi \right) \right\rangle,
\label{op}
\end{equation}
where $\langle \cdots \rangle$ is the time average taken
after achieving the steady state and
$\Delta \phi \equiv  |\phi_1 - \phi_2|$
with the phase $\phi_i$ determined from $\theta_i \propto \sin\phi_i$.
For this, we neglect the result generated in the initial duration of $T_{\rm ini} = 10^5$, and then take the average during the time $T = 10^4$.
Our definition of $m$ gives us $m=1$ if the two pendulums keep
the antiphase synchronization, while $m=-1$ is obtained for
the complete inphase synchronization.  The smaller $|m|$ is, the worse
the synchronization occurs.
In addition to the phase synchronization of oscillators,
a related but distinct phenomenon is the frequency entrainment~\cite{freq}.
We also gauge the degree of the frequency entrainment simply by measuring the
frequency difference defined by
\begin{equation}
\Delta f \equiv \left | \frac{N_1}{T} - \frac{N_2}{T} \right |,
\label{fd-syn}
\end{equation}
where $N_i$ is the number of oscillations of the pendulum $i$
during the time $T$ after the initial transient period $T_{\rm ini}$.

In the numerical computation, we change $\bar l_2$ from $1$ to $1.006$ with the
interval $1.2\times 10^{-4}$, for fixed $\bar l_1 = 1$~\cite{inherentdelta}. For each pair of $(\bar l_1, \bar l_2)$ prepared in this way, we assign the quenched random number
$\theta_i \in \pm[0.05, 0.1]$ at the time $t=0$ and also 
$x= {\dot x} = \dot\theta_1 = \dot\theta_2 = 0$ is used as the initial condition.
We then measure the two key quantities $m$ and $\Delta f$
at given values of the length difference
\begin{equation}
\Delta {\bar l} \equiv | {\bar l}_1 - {\bar l}_2 | .
\end{equation}
When $\Delta {\bar l} = 0$, the two pendulums have identical natural frequency
and we expect them to show perfect antiphase synchronization~\cite{Bennett}
to give us $m = 1$ and $\Delta f = 0$. As the length difference
becomes larger, it is expected that beyond some value of $\Delta {\bar l}$ the
system should stop showing synchronization resulting in $m < 1$ and $\Delta f \neq 0$.

Figure~\ref{bistability} summarizes our main results: As is expected, one can
clearly see the existence of dynamic phase transition at
$\Delta {\bar l}_c \approx 0.002$ which splits the
antiphase synchronized state ($m \approx 1$ and $\Delta f \approx 0$)
and the desynchronized state ($m < 1$ and $\Delta f > 0$) as the length
difference $\Delta {\bar l}$ is increased. The measured two order
parameters are also shown not to have intermediate values around the
dynamic phase transition, indicating the discontinuous nature of the
transition. Another interesting observation is the existence of the
broad range of bistability: Whether or not the system approaches
the synchronized state is not uniquely determined by the length
difference $\Delta {\bar l}$ only.
Although not shown here, we also observe that the use of the
larger reduced mass $\mu$ increases $\Delta {\bar l}_c$, which has also been
reported in Ref.~\onlinecite{Bennett}.

\begin{figure}
\includegraphics[width=0.45\textwidth]{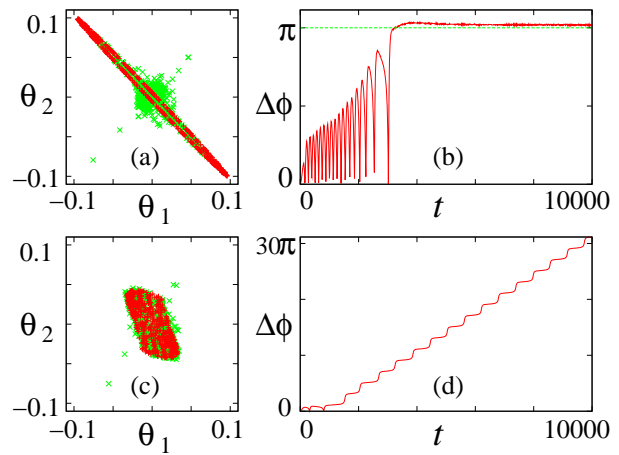}
\caption{(Color online) The representative details of pendulum motions are depicted for
the antiphase synchronized state [$\Delta {\bar l} = 9.6\times 10^{-4}$, (a) and (b)]
and for the desynchronized state [$\Delta {\bar l} = 39.6\times10^{-4}$, (c) and (d)] for $\mu = 0.025$.
In (a) and (c), the time evolution of angular coordinates is exhibited
in the plane of ($\theta_1$,$\theta_2$), green for $t \leq 5000$ and
red for $t > 5000$, and in (b) and (d) the phase
difference $\Delta \phi(t)$ is shown as a function of the time $t$.
In the antiphase synchronized state, (a) $\theta_1$ and $\theta_2$ eventually
align along the line with the negative slope in the plane
($\theta_1, \theta_2$) since (b) $\Delta \phi \rightarrow \pi$.
The desynchronized state is characterized by (c) the scattered points
in the $(\theta_1, \theta_2)$ plane, and (d) the indefinite increase of
$\Delta \phi$ as $t$ is increased due to the mismatch of the frequencies
($\Delta f \neq 0$).
}
\label{detail}
\end{figure}

We next investigate in Fig.~\ref{detail} the features of the antiphase
synchronized state [(a) and (b) for $\Delta {\bar l} = 9.6\times 10^{-4}$] 
and the desynchronized state [(c) and (d) for $\Delta {\bar l} = 39.6\times 10{-4}$]
for the reduced mass $\mu = 0.025$.
In Fig.~\ref{detail}(a) for the antiphase synchronized state,
one finds that there is a manifest anticorrelation between $\theta_1$
and $\theta_2$ due to the phase difference $\Delta \phi = \pi$ after
some initial transient period.
The approach toward the antiphase synchronization is clearly displayed
in Fig.~\ref{detail}(b): Again after transients, $\Delta \phi$ 
approaches the odd multiple of $\pi$.  In contrast, the desynchronized state shown
in Fig.~\ref{detail}(c) and (d) exhibit very different behaviors:
The anticorrelation between $\theta_1$ and $\theta_2$ becomes much
weaker [see (c)]
and the phase difference $\Delta \phi$ increases indefinitely
in time [see (d)], due to the nonzero frequency difference $\Delta f \neq 0$
[see Fig.~\ref{bistability}(b)].

\begin{figure}
\includegraphics[width=0.48\textwidth]{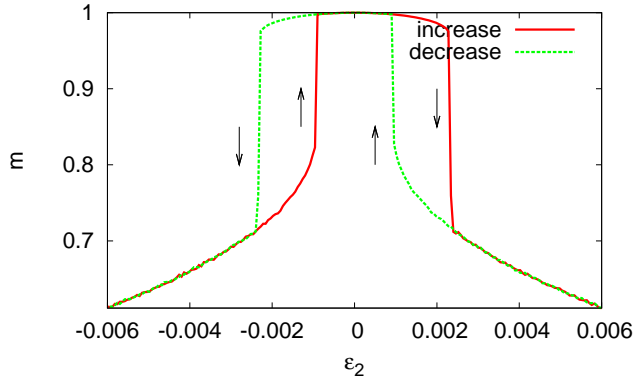}
\caption{(Color online) The phase synchronization order parameter $m$ is
measured as the length ${\bar l_2}$ of the second pendulum is
increased (the full red curve) or decreased (the dotted green curve) adiabatically,
while the length of the first pendulum is fixed to ${\bar l}_1 = 1$,
and ${\bar l}_2 = 1 + \epsilon_2$.
The hysteresis effect is clearly seen around the
dynamic phase transition splitting the antiphase synchronized (the upper branch
of the curves) state and the desynchronized state (the lower branch).
}
\label{hys}
\end{figure}

We finally examine that there is a hysteresis effect around the dynamic phase
transition between the antiphase synchronized phase and the desynchronized
phase. For this, we fix the length of the first pendulum to ${\bar l}_1 = 1$
and change adiabatically ${\bar l}_2$.  For the given value of
${\bar l}_2$, we integrate equations of motion for a sufficiently long time to
achieve the steady state and change ${\bar l}_2$ by $6.0\times 10^{-5}$. Note that since the hysteresis is of interest now, the system is not reinitialized after the control parameter ${\bar l}_2$ is
changed. As clearly shown in Fig.~\ref{hys}, our two pendulum system
manifests the hysteresis behavior around the dynamic phase transition:
The phase synchronization order parameter
$m$ follows different curves when the length of the second pendulum
is increased (the red curve in Fig.~\ref{hys}) and decreased (the green one
in Fig.~\ref{hys}).
%We note that the bistable region is wider in Fig.~\ref{hys}
%than observed in Fig.~\ref{bistability}(a).
All the observations, i.e.,
the discontinuous nature of the dynamic phase transition,
the broad range of bistability, and the strong hysteresis effect,
strongly suggest that a subcritical Hopf bifurcation takes place there
if stated in the language of nonlinear dynamics~\cite{Strogatz}.
%can be understood as the results from the
%subcritical Hopf bifurcation if stated in the language of
%nonlinear dynamics~\cite{Strogatz}.

In summary, we have numerically studied the synchronization of the two
nonidentical pendulum motions, pivoting on a common movable frame.
Within the limitation of our setup of numerical experiments,
it has been clearly shown that the system exhibits the discontinuous dynamic
phase transition from the antiphase synchronized state to the
desynchronized state as the length difference is increased.
We have also shown that the discontinuous nature of the transition is
reflected to the broad range of bistability and also to the existence
of the hysteresis effect.  We believe that our model reproduces
the odd sympathy C. Huygens observed in the seventeenth
century.
We finally remark that it is well-known that the dynamic property of nonlinear
system is significantly affected by such mechanical details if involved in the
nonlinearity~\cite{Jackson}. In the present system, it is the escapement
mechanisms~\cite{Bennett,Pantaleon} or the structural design of the
system~\cite{Senator,Alexander}, for example. Therefore, it is still necessary
to examine the various nonlinearity setups for the sound understanding
of the synchronization phenomenon in general as a dynamic phase in statistical
physics.

This research was supported by 
WCU(World Class University) program through the National Research Foundation of Korea funded by the Ministry of Education, Science and Technology  (R31-2008-000-10029-0).


\begin{thebibliography}{99}
\bibitem{inherentdelta} We consider $\Delta \bar l$ not only to examine
when the synchronization begins but also to reflect the generic
inhomogeneity of the real pendulums. The main interest of 
synchronization phenomena is indeed on the transition point where 
the dynamic process overcomes the intrinsic difference, arriving at 
the sympathy. In the simple setup prsented here, the difference of the
natural frequencies of the two pendulums is surely the first candidate 
to prohibit them from synchronizing to each other. We
consider $\Delta \bar l$ to effectively introduce the natural frequency
difference, playing the role of the variance in the frequency distribution
in the Kuramoto model (see Ref.~\onlinecite{freq} and references therein for example).
\end{thebibliography}
\end{document}